\documentclass[%
 reprint,
 amsmath,amssymb,
 aps,prc
]{revtex4-1}

\usepackage{graphicx}
\usepackage{dcolumn}
\usepackage{bm}
\usepackage{hyperref}
\usepackage{amsmath}

\newcommand{\Ham}{\mathcal{H}}

\begin{document}

\preprint{APS/123-QED}

\title{Isovector properties of the nuclear energy density functional from the quark-meson coupling model}

\author{E. McRae}
\email{ellen.mcrae@anu.edu.au}
\affiliation{Department of Nuclear Physics, Research School of Physics and Engineering, The Australian National University, Canberra, ACT 2601, Australia}

\author{C. Simenel}
\email{cedric.simenel@anu.edu.au}
\affiliation{Department of Nuclear Physics, Research School of Physics and Engineering, The Australian National University, Canberra, ACT 2601, Australia}

\author{E. C. Simpson}
\affiliation{Department of Nuclear Physics, Research School of Physics and Engineering, The Australian National University, Canberra, ACT 2601, Australia}

\author{A. W. Thomas}
\affiliation{ARC Centre of Excellence for Particle Physics at the Terascale and CSSM, Department of Physics, The University of Adelaide, SA 5005, Australia}

\date{\today}

\begin{abstract}

\begin{description}

\item[Background]

The Skyrme energy density functional is widely used in mean-field calculations of nuclear structure and dynamics. However, its reliance on phenomenology may compromise its isovector properties and its performance for exotic nuclear systems.

\item[Purpose]

This work investigates the possibility of removing some phenomenology from the density functional by drawing on the high-energy degrees-of-freedom of the quark-meson coupling (QMC) model. The QMC model has microscopically derived isovector properties and far fewer adjustable parameters.

\item[Method]

The parameters of the Skyrme interaction are fixed using the energy density functional of the QMC model, to give the Skyrme-QMC (SQMC) parameterisation.

\item[Results]

Hartree-Fock-Bogoliubov calculations of the Sn, Pb and $N=126$ chains are reported, in which SQMC performs with an accuracy comparable to modern phenomenological functionals.

\item[Conclusions]

The isovector terms of the energy density functional are significant for the binding energies of neutron-rich nuclei. The isovector dependence of the nuclear spin-orbit interaction must be taken into account for calculations of r-process nucleosynthesis abundances.

\end{description}

\end{abstract}

\maketitle

\section{\label{sec:Intro}Introduction}
Interest in exotic nuclear systems is growing as new experimental facilities and techniques constantly expand the limits of known nuclei. Astrophysical models, of r-process nucleosynthesis \cite{LangankeWiescher} and neutron stars for example, rely heavily on the properties of very exotic nuclei, many of which are unknown. To guide these searches predictive theories of nuclear physics are required.
\par
However, one must face the significant challenges posed by the quantum many-body problem. This is further complicated by the interaction between nucleons, which is not well understood in-medium. For very neutron-rich nuclei the isovector properties of the interaction, which are far less known than the isoscalar channels, will become increasingly important.
\par
One possible approach is to consider systematic applications of chiral effective field theory \cite{Weinberg1991} to nuclear interactions and hence to nuclear structure. 
It is certainly appealing to develop such an approach based on the symmetries of quantum chromodynamics. 
Nevertheless, chiral perturbation theory does not yet allow systematic applications to structure and dynamics across the nuclear chart.
\par
Self-consistent mean-field approaches based  on  nuclear energy density functionals (EDF) make the problem computationally tractable. The EDF approach relies on the Hohenberg and Kohn theorem~\cite{Hohenburg1964}, according to which the nucleon density distributions uniquely determine the ground-state energy of the nuclear system.
\par
An energy density functional can be derived from an effective nucleon-nucleon force, for example the Skyrme interaction, and used at the mean-field level in a range of approaches based on Hartree-Fock approximations, which offer a unified description of nuclear structure and dynamics~\cite{Negele1982,Bender2003,Simenel2012,nakatsukasa2016}. However, this requires the introduction of some amount of phenomenology. For instance, the form of the density dependence of the coupling constants is essentially guessed. As a result, the Skyrme EDF contains a large number of parameters without clear physical meanings, compromising its predictive power when applied to very exotic systems.
\par
Another difficulty in the Skyrme approach is  that the strength of the spin-orbit coupling must be fitted to experimental data, such as the energy splitting between spin-orbit single-particle level partners. 
Such data are often not known away from stability.
This leads to difficulties in determining the isovector dependence of the spin-orbit force.
It is thus desirable to develop approaches which would account for the  spin-orbit interaction without requiring additional parameters. 
Indeed, including as little phenomenology as possible should improve the reliability of predictions for very exotic nuclei. 
\par
In fact, few models can predict the nuclear spin-orbit interaction, a relativistic effect which, unlike the electronic spin-orbit force, is not a small correction. As shown by Goeppert-Mayer and Jensen \cite{GoeppertMayer}, the inclusion of the nuclear spin-orbit coupling is essential for correctly predicting all magic numbers above 20.
The strength of the spin-orbit interaction plays a role in determining the location of the superheavy island of stability \cite{bender1999}, as well as providing a dissipation mechanism which is crucial in heavy-ion collisions~\cite{umar1986}.
\par
Covariant energy density functionals  offer another interesting approach to the nuclear many-body problem \cite{ring1996}.
Being fully relativistic, they naturally generate a spin-orbit interaction and, if exchange terms are properly included, some isovector dependence \cite{ebran2016}. 
However, as in the Skyrme EDF, the density dependence of the coupling constants requires some phenomenology, leading to a relatively large number of parameters needed to fit experimental data or pseudodata. 
\par
One promising avenue is to obtain the nuclear EDF by drawing on the high-energy degrees-of-freedom of the nuclear system through a self-consistent mean-field model of the in-medium modification of the quark structure of the bound nucleons: the quark-meson coupling (QMC) model~\cite{Guichon1996}. The QMC model predicts both the central and spin-orbit channels of the effective nucleon-nucleon interaction, along with their isovector behaviour. This is achieved with a much smaller number of parameters than in standard Skyrme or covariant EDF approaches.
\par
There exist many codes based on Hartree-Fock approximations and their generalisations that can be used to compute the properties of atomic nuclei and their reactions (see, e.g., \cite{HFBRAD,schunck2012,HFBTHO,maruhn2014,ryssens2015}). The majority of these codes are based on Skyrme functionals. To take advantage of the Skyrme framework, this work looks at the process of using the QMC model to fix the parameters of a Skyrme functional. With applications to exotic nuclei in mind, we focus on the importance of the isovector behaviour of the energy density functional and, in particular, the isovector dependence of the spin-orbit term as derived within the QMC model.
\par
The philosophy of the QMC model is briefly outlined in Sec.~\ref{sec:QMC}. Section~\ref{sec:Skyrme} introduces the Skyrme energy density functional. A method for fixing the Skyrme parameters from the QMC model is then detailed in Sec.~\ref{sec:SQMC}. Hartree-Fock-Bogoliubov (HFB) calculations are presented in Sec.~\ref{sec:Results}, along with an analysis of the level of success of this method. Finally, the spin-orbit functional and its importance for exotic nuclei is investigated in Sec.~\ref{sec:Spin-orbit}, including a comparison to the approach of standard relativistic mean-field (RMF) theory.

\section{\label{sec:QMC}The quark-meson coupling model}
The quark-meson coupling model is a relativistic mean-field approach to the quantum many-body problem. Proposed by P. A. M. Guichon in Ref.~\cite{Guichon1988} and since developed in Refs.~\cite{Guichon1996,Guichon2006,Stone2016}, the QMC model posits that the nucleon is not immutable and that modification of the internal quark structure by the nuclear environment is an important factor in modelling nuclear systems. In its simplest form, the model considers three valence quarks confined to an MIT bag \cite{MITbag}, which interact with the quarks of the surrounding nucleons via the exchange of $\sigma$, $\omega$ and $\rho$ mesons. The $\omega$ and $\rho$ fields are identified with the real vector-isoscalar and vector-isovector particles, respectively. The $\sigma$ field is an effective representation of scalar-isoscalar exchange, including two correlated pions.
\par
Thanks to its inclusion of quark degrees-of-freedom, the QMC model can be used to describe effects associated with the modification of the structure of hadrons in the nuclear medium. 
As an example, it proposes a possible explanation \cite{Thomas1989} to the famous ``European Muon Collaboration'' (EMC) effect, which involves the loss of momentum from valence quarks in a nucleus compared with a free nucleon. (Note that an alternate development by Bentz and Thomas has applied the same concept to the Nambu-Jona-Lasinio (NJL) model~\cite{Bentz:2001vc}, which allows a covariant formulation of the calculation of nuclear structure functions, with similar results~\cite{Cloet:2006bq,Cloet:2012td}.)
The QMC model was also applied to the masses and properties of other hadrons immersed in a nuclear medium~\cite{Saito:2005rv}, which amongst other things led to a remarkably successful description of hypernuclei \cite{Tsushima1998,Guichon:2008zz}. In particular, the QMC model naturally explains the virtual absence of a spin-orbit force in $\Lambda$ hypernuclei, repulsion for $\Sigma$ hyperons in nuclei and it predicted binding at the level of a few MeV for $\Xi$ hypernuclei, for which the first evidence has recently been reported at  JPARC  \cite{Nakazawa2015}.
\par
Beginning from a Lorentz-invariant Lagrangian density \cite{Guichon1996} of quark and meson fields, the system is solved self-consistently to give a non-relativistic energy density functional~\cite{Guichon2006} which can be applied to the study of finite nuclei. The EDF can be written as
\begin{equation}
\Ham=\rho M_{N}+\frac{\tau}{2M_{N}}+\Ham_{0+3}+\Ham_{\textrm{eff}}+\Ham_{\textrm{fin}}+\Ham_{\textrm{SO}},
\label{eq:EDF}
\end{equation}
separated based on the occurrence of the total (neutron, proton) particle density, $\rho$ ($\rho_{n}$, $\rho_{p}$); kinetic density, $\tau$ ($\tau_{n}$, $\tau_{p}$); and spin-orbit density, $\bm{J}$ ($\bm{J}_{n}$, $\bm{J}_{p}$). The QMC EDF of Ref.~\cite{Guichon2006} gives
\begin{widetext}
\begin{eqnarray}
\Ham_{0+3}^{\textrm{QMC}}=&&\left(\frac{-3G_{\rho}}{32}+\frac{G_{\sigma}}{8\left(1+d\rho G_{\sigma}\right)^{3}}-\frac{G_{\sigma}}{2\left(1+d\rho G_{\sigma}\right)}+\frac{3G_{\omega}}{8}\right)\rho^{2} 
+\left(\frac{5G_{\rho}}{32}+\frac{G_{\sigma}}{8\left(1+d\rho G_{\sigma}\right)^{3}}-\frac{G_{\omega}}{8}\right)\left(\rho_{n}-\rho_{p}\right)^{2}, \label{eq:H0H3QMC} \\
\Ham_{\textrm{eff}}^{\textrm{QMC}}=&&\left(\frac{G_{\rho}}{4m_{\rho}^{2}}+\frac{G_{\sigma}}{2M_{N}^{2}}\right)\rho\tau+\left(\frac{-G_{\rho}}{8m_{\rho}^{2}}-\frac{G_{\sigma}}{2m_{\sigma}^{2}}+\frac{G_{\omega}}{2m_{\omega}^{2}}-\frac{G_{\sigma}}{4M_{N}^{2}}\right)\sum_{q=n,p}\rho_{q}\tau_{q}, \label{eq:HeffQMC}\\
\Ham_{\textrm{fin}}^{\textrm{QMC}}=&&\left(\frac{-3G_{\rho}}{16m_{\rho}^{2}}-\frac{G_{\sigma}}{2m_{\sigma}^{2}}+\frac{G_{\omega}}{2m_{\omega}^{2}}-\frac{G_{\sigma}}{4M_{N}^{2}}\right)\rho\nabla^{2}\rho 
+\left(\frac{9G_{\rho}}{32m_{\rho}^{2}}+\frac{G_{\sigma}}{8m_{\sigma}^{2}}-\frac{G_{\omega}}{8m_{\omega}^{2}}+\frac{G_{\sigma}}{8M_{N}^{2}}\right)\sum_{q=n,p}\rho_{q}\nabla^{2}\rho_{q}\textrm{, and}\label{eq:HfinQMC}\\
\Ham_{\textrm{SO}}^{\textrm{QMC}}=&&-\frac{1}{4M_{N}^{2}}\left[\left(G_{\sigma}+G_{\omega}\left(2\mu_{s}-1\right)\right)\rho\nabla\cdot\bm{J}
+\left(\frac{G_{\sigma}}{2}+\frac{G_{\omega}}{2}\left(2\mu_{s}-1\right)+\frac{3G_{\rho}}{8}\left(2\mu_{v}-1\right)\right)\sum_{q=n,p}\rho_{q}\nabla\cdot\bm{J}_{q}\right].\label{eq:HsoQMC}
\end{eqnarray}
\end{widetext}
One assumes experimental values for the free nucleon mass, $M_{N}$, the $\omega$ and $\rho$ meson masses, $m_{\omega}$ and $m_{\rho}$, and the isoscalar and isovector nucleon magnetic moments, $\mu_{s}$ and $\mu_{v}$. This leaves only the $\sigma$ meson mass, $m_{\sigma}$, and a coupling constant for each of the three meson fields, $G_{\sigma}$, $G_{\omega}$ and $G_{\rho}$, as free parameters in the model.  The `scalar polarisability', $d$, is a constant which describes the self-consistent response of the confined quarks to the applied scalar mean-field. It gives rise very naturally to the many-body interactions within the system of nucleons, as we see from the appearance of $d$ in the denominators of Eq.~(\ref{eq:H0H3QMC}).
\par
In this work, as in Ref.~\cite{Guichon2006}, for an appropriate choice of $m_{\sigma}$ (e.g. 700 MeV), the meson coupling constants are fixed using infinite nuclear matter pseudodata: the nucleon density, energy per nucleon, and symmetry energy at saturation ($\rho_{0}=0.16$ fm$^{-3}$, $e_{\infty}=-15.85$ MeV, and $a_{s}=30$ MeV, respectively).
As the parameters are not determined by a fit to experimental masses and radii, the functional can be used consistently in both the intrinsic frame of a nucleus and the centre-of-mass frame of a collision between two nuclei \cite{Simenel2012}.
Reference~\cite{Stone2016} on the other hand, fixes the free parameters of the QMC model through a fit to ground-state properties of nuclei, while ensuring consistency with nuclear matter properties within reasonable uncertainties, obtaining similar values.

\section{\label{sec:Skyrme}The Skyrme functional}
The Skyrme interaction began as a general form dictated by the symmetries that are required of the force between nucleons \cite{Skyrme1959}. It has since been generalised to an energy density functional of local densities and currents up to second-order in derivatives. Separated into the terms of Eq.~(\ref{eq:EDF}), the Skyrme EDF takes the form:
\begin{widetext}
\begin{eqnarray}
\Ham_{0+3}^{\textrm{Skyrme}}=&&\frac{1}{12}\left\{\left[6t_{0}\left(1+\frac{x_{0}}{2}\right) + t_{3}\left(1+\frac{x_{3}}{2}\right)\rho^{\alpha}\right] \rho^{2} -\left[6t_{0}\left(x_{0}+\frac{1}{2}\right)+t_{3}\left(x_{3}+\frac{1}{2}\right)\rho^{\alpha}\right] \sum_{q=n,p}\rho_{q}^{2}\right\}, \label{eq:H0H3Skyrme} \\
\Ham_{\textrm{eff}}^{\textrm{Skyrme}}=&&\frac{1}{4}\left\{\left[t_{1}\left(1+\frac{x_{1}}{2}\right)+t_{2}\left(1+\frac{x_{2}}{2}\right)\right]\rho\tau - \left[t_{1}\left(x_{1}+\frac{1}{2}\right)-t_{2}\left(x_{2}+\frac{1}{2}\right)\right]\sum_{q=n,p}\rho_{q}\tau_{q}\right\}, \label{eq:HeffSkyrme}\\
\Ham_{\textrm{fin}}^{\textrm{Skyrme}}=&&-\frac{1}{16}\left\{\left[3t_{1}\left(1+\frac{x_{1}}{2}\right)-t_{2}\left(1+\frac{x_{2}}{2}\right)\right]\rho\nabla^{2}\rho - \left[3t_{1}\left(x_{1}+\frac{1}{2}\right)+t_{2}\left(x_{2}+\frac{1}{2}\right)\right]\sum_{q=n,p}\rho_{q}\nabla^{2}\rho_{q}\right\}\textrm{, and}\label{eq:HfinSkyrme}\\
\Ham_{\textrm{SO}}^{\textrm{Skyrme}}=&&-\frac{1}{2}\left(W_{0}\rho\nabla\cdot\bm{J} + W_{0}'\sum_{q=n,p}\rho_{q}\nabla\cdot\bm{J}_{q}\right) , \label{eq:HsoSkyrme}
\end{eqnarray}
\end{widetext}
with eleven  parameters $x_{0-3}$, $t_{0-3}$, $\alpha$, $W_{0}$, and $W_{0}'$. 
These parameters are typically determined by a fit to nuclear equation of state properties and to the experimental masses and radii of a selection of atomic nuclei~\cite{Bender2003}. 
Most modern functionals also include uncertainties on their parameters~\cite{Kortelainen2010, Kortelainen2012,kortelainen2014}.
\par
While Skyrme functionals remain a widely successful approach, the phenomenological nature of the traditional Skyrme functional can call into question its predictions for very exotic systems. Because of the correlations between parameters and the many choices of fitting procedure, which place an  emphasis on different experimental data, many Skyrme parameterisations exist. They give conflicting predictions for the properties of very exotic nuclei \cite{Mumpower2016}, with no clear winner.

\section{\label{sec:SQMC}The Skyrme-QMC functional}
Despite its limitations, the Skyrme functional is very convenient and the vast majority of Hartree-Fock codes are based on it. In an attempt to improve the predictive power of the Skyrme-Hartree-Fock approach, this paper determines the parameters of the Skyrme EDF using the QMC model. 
In essence, this is achieved by equating the terms of the QMC density functional [Eqs.~(\ref{eq:H0H3QMC}~--~\ref{eq:HsoQMC})] to those of the Skyrme functional [Eqs.~(\ref{eq:H0H3Skyrme}~--~\ref{eq:HsoSkyrme})]. 

\subsection{QMC700}
A similar approach was adopted by Guichon \textit{et al.} \cite{Guichon2006} to produce the QMC700 Skyrme parameterisation (``700'' refers to the choice of $m_\sigma=700$~MeV).
QMC700 was obtained by approximating the QMC functional [Eqs.~(\ref{eq:H0H3QMC}~--~\ref{eq:HsoQMC})] with a simplified Skyrme EDF analogous to the SkM* parameterisation \cite{SkM*}.
As a result,  QMC700 had only 6 parameters ($t_{0-3}$, $x_{0}$, and $W_{0}$), setting $x_{1-3}=0$, $W_{0}=W_{0}'$, and $\alpha=1/6$ (see Table~\ref{tab:Parameters}). Because of  the reduced number of parameters, the $\Ham_{\textrm{eff}}$, $\Ham_{\textrm{fin}}$ and $\Ham_{\textrm{SO}}$ functionals of Skyrme and QMC were equated for only $N=Z$ (assuming $\rho_{n}=\rho_{p}$), to yield $t_{1}$, $t_{2}$ and $W_{0}$. The microscopically derived density dependence of the $\Ham_{0+3}$ term of the QMC functional [Eq.~(\ref{eq:H0H3QMC})] is significantly more complicated than that of the Skyrme functional [Eq.~(\ref{eq:H0H3Skyrme})] because of the terms involving the scalar polarisability~$d$. Unlike the other terms, $\Ham_{0+3}^{\textrm{Skyrme}}$ cannot be equated to $\Ham_{0+3}^{\textrm{QMC}}$. Instead it is fitted numerically  over a large range of nucleon density, $\rho\in[0, 0.2\textrm{ fm}^{-3}]$ with $\rho_n=\rho_p$, to give $t_{0}$ and $t_{3}$, and with $Z/A=82/208$ (corresponding to $^{208}$Pb) to give $x_{0}$~\cite{Guichon2006}. Out of 240 Skyrme parameterisations tested in Ref.~\cite{Dutra}, QMC700 and QMC650 (a parameterisation derived identically but with $m_{\sigma}=650$ MeV) were two of only 16 parameter sets shown to satisfy all constraints (a range of nuclear matter properties derived from experiment).
\begin{table*}
\caption{\label{tab:Parameters}Skyrme parameters.}
\begin{ruledtabular}
\begin{tabular}{lllllll}
                                   &QMC700 \cite{Guichon2006}&QMC700*  & SQMC     &SQMC$^{\textrm{(Wang)}}$ \cite{Wang2011}&SQMC$^{\alpha_{1},\alpha_{2}}$& UNEDF1 \cite{Kortelainen2012}\\
\colrule
$t_{0}$ (MeV fm$^{3}$)             &-2429.1 	             &-2643.869& -2643.869& -2648.19                               & -843.3117                     &-2078.3280\\
$t_{1}$ (MeV fm$^{5}$)             & 370.975 	             &370.829  & 370.829  &  371.07                                & 370.829                       & 239.40081\\
$t_{2}$ (MeV fm$^{5}$)             & -96.6902	             & -99.5702& -121.666 & -121.644                               & -121.666                      & 1575.1195\\
$t_{3}$ (MeV fm$^{3+3\alpha_{1}}$) & 13773.4 	             & 15490.99& 15490.99 &  15553.495                             & -8166.466                     & 14263.646\\
$x_{0}$                            & 0.1                     & 0.093084& 0.59761  &  0.60146                               & 1.74627                       & 0.0537569\\
$x_{1}$                            & 0 	                     & 0       & 0.266456 &  0.2697                                & 0.266456                      & -5.077232\\
$x_{2}$                            & 0 	                     & 0       & -0.227015&  -0.23701                              & -0.227015                     & -1.366506\\
$x_{3}$                            & 0 	                     & 0       & 0.693348 &  0.6968                                & -2.372489                     & -0.162491\\
$\alpha_{1}$                       & $1/6$                   & $1/6$   & $1/6$    &  $1/6$                                 & $1/3$                         &  0.270018\\
$W_{0}$ (MeV fm$^{5}$)             & 104.58                  & 104.3872& 82.8602  &  104.498                               & 82.8602                       & 76.736144\\
$W_{0}'$ (MeV fm$^{5}$)            & 104.58 	             & 104.3872& 147.4411 &  104.498                               & 147.4411                      & 142.63304\\
$t_{6}$ (MeV fm$^{3+3\alpha_{2}}$) & 0                       & 0       & 0        &  0                                     & 17069.2                       & 0    \\
$x_{6}$                            & 0                       & 0       & 0        &  0                                     & -0.645158                     & 0    \\
$\alpha_{2}$                       & 0                       & 0       & 0        &  0                                     & $2/3$                         & 0    \\
\end{tabular}
\end{ruledtabular}
\end{table*}
\subsection{QMC700*}
\begin{figure}
\includegraphics[width=\columnwidth]{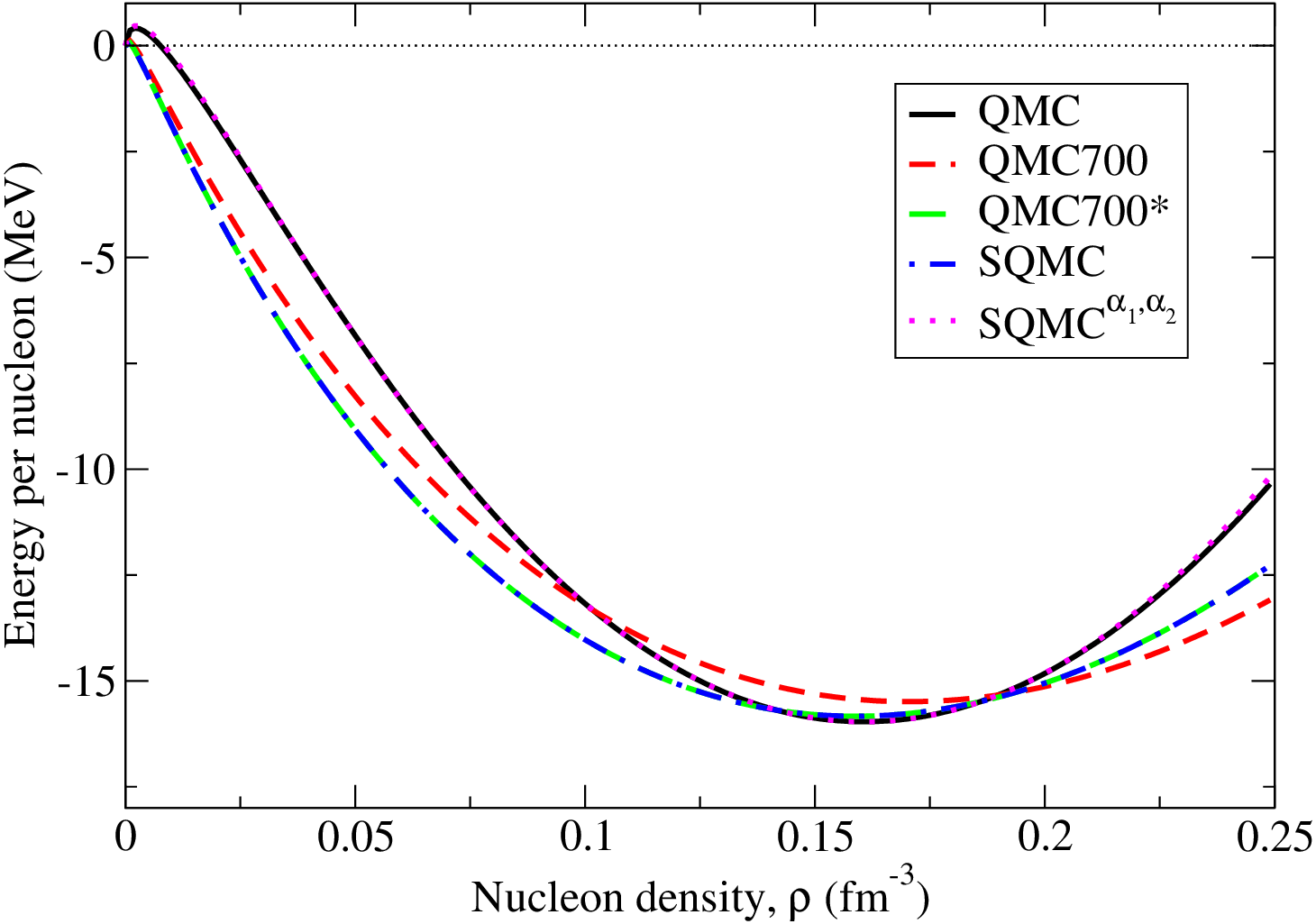}
\caption{\label{fig:EOS}Symmetric nuclear matter equations of state.}
\end{figure}

The complicated density dependence of the QMC model [Eq.~(\ref{eq:H0H3QMC})] cannot be well reproduced by a Skyrme functional with a single $\rho^\alpha$ dependence as in Eq.~(\ref{eq:H0H3Skyrme}). 
This is illustrated in Fig.~\ref{fig:EOS}, which shows the equations of state for symmetric nuclear matter for various EDF parameterisations.
By restricting the range of density used in the $\Ham_{0+3}$ fit to $\rho\in[0.12, 0.2\textrm{ fm}^{-3}]$, the reproduction of the QMC functional is improved around saturation density where this bulk term will be most important (see also Table~\ref{tab:INMproperties}).
Hartree-Fock calculations of nuclear masses and radii show a corresponding improvement in performance as compared to QMC700. This refitting alters $t_{0}$, $t_{3}$ and $x_{0}$, and we label this new parameter set QMC700*. The other parameters remain unchanged, so that QMC700* has the same isovector properties as QMC700.
\begin{table*}
\caption{\label{tab:INMproperties}Properties of symmetric nuclear matter: saturation density $\rho_{0}$, energy per nucleon at saturation $e_{\infty}$, incompressibility $K_{\infty}$, isoscalar effective nucleon mass $m^{*}_{s}$, enhancement factor of the Thomas-Reiche-Kuhn sum rule $\kappa$ (related to the isovector effective mass), and symmetry energy $a_{s}$.}
\begin{ruledtabular}
\begin{tabular}{lllllll}
                               & $\rho_0$ (fm$^{-3}$) & $e_{\infty}$ (MeV) & $K_{\infty}$ (MeV) & $m^{*}_{s}/M_{N}$ & $\kappa$& $a_{s}$ (MeV)\\
\colrule
 QMC                           & 0.1607               & -15.95             & 347.5              & 0.7706            & 0.6050  & 29.93 \\
 QMC700                        & 0.1704               & -15.49             & 219.6              & 0.7556            & 0.5636  & 33.40 \\
 QMC700*                       & 0.1588               & -15.84             & 218.7              & 0.7727            & 0.5193  & 32.09 \\
 SQMC                          & 0.1588               & -15.84             & 218.7              & 0.7727            & 0.5980  & 29.87 \\
 SQMC$^{\alpha_{1},\alpha_{2}}$& 0.1609               & -15.97             & 350.7              & 0.7704            & 0.6059  & 29.97 \\ 
\end{tabular}
\end{ruledtabular}
\end{table*}
\subsection{SQMC}
Given that our aim is to study the isovector dependence of the nuclear density functional and its importance for exotic systems, we create one final QMC-motivated Skyrme force by extending QMC700* to include additional isovector parameters, $x_{1-3}$ and $W_{0}'$. This yields a Skyrme-QMC functional which perfectly reproduces $\Ham_{\textrm{eff}}^{\textrm{QMC}}$, $\Ham_{\textrm{fin}}^{\textrm{QMC}}$ and $\Ham_{\textrm{SO}}^{\textrm{QMC}}$ for all densities and nucleon asymmetries, and improves the fit for $\Ham_{0+3}^{\textrm{QMC}}$. This parameterisation is labelled ``SQMC'' and is the primary focus of this paper. It is consistent with the parameterisation of Ref.~\cite{Wang2011} (labelled SQMC$^{\mbox{(Wang)}}$ in the following) for the central terms. 
However, SQMC$^{\mbox{(Wang)}}$ does not include the second spin-orbit parameter, fixing $W_{0}=W_{0}'$, unlike the present version of SQMC.
\subsection{Incompressibility}
The incompressibility of symmetric nuclear matter has long been the subject of debate. The QMC model parameterisation in Ref.~\cite{Guichon2006} has an incompressibility of 346 MeV, in line with most other relativistic approaches. For all Skyrme parameterisations discussed thus far (QMC700, QMC700* and SQMC), the density dependence is fixed with $\alpha=1/6$. This restricts the functionals to a lower incompressibility of around 220 MeV, consistent with many common Skyrme functionals, such as SLy4~\cite{chabanat1998} and UNEDF1~\cite{Kortelainen2012}.
\par
\begin{figure}
\includegraphics[width=\columnwidth]{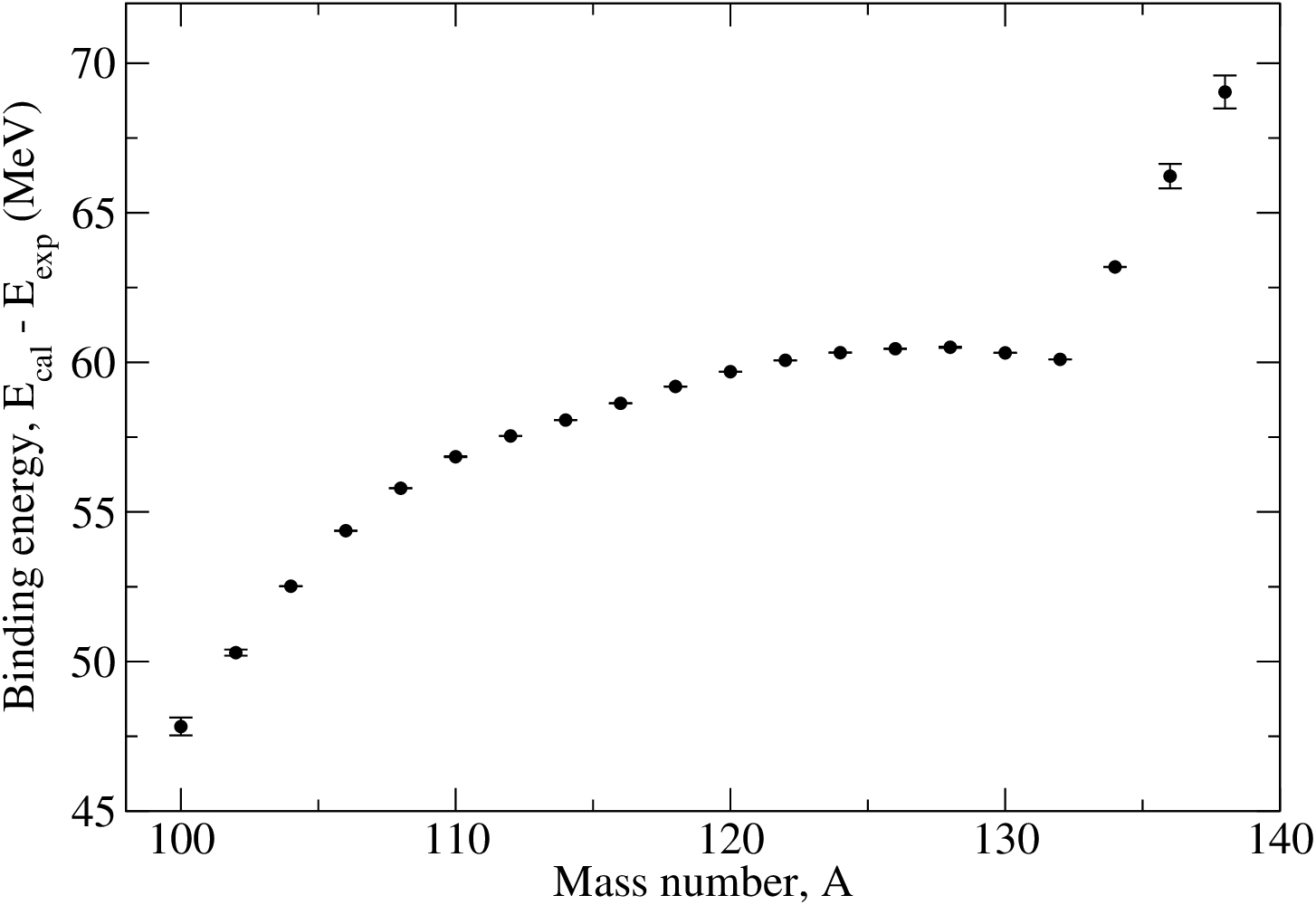}
\caption{\label{fig:SQMC4_Sn}Difference between calculated and experimental
  ground-state binding energies for $^{A}$Sn nuclei using the
  SQMC$^{\alpha_1,\alpha_2}$ parameterisation. The error bars are from the
  experimental data $(E_{\textrm{exp}})$ \cite{AME2012}.}
\end{figure}

To generate a QMC functional with an incompressibility closer to the QMC model, we use the spherical Hartree-Fock-Bogoliubov code \textsc{hfbrad}~\cite{HFBRAD}, which allows for two fractional density dependences in $\Ham_{0+3}$. This makes it possible to reproduce the full QMC functional very accurately over a wide range of densities, using $\alpha_{1}=1/3$ and $\alpha_{2}=2/3$, as shown in Fig.~\ref{fig:EOS}. The resulting Skyrme parameterisation, labelled SQMC$^{\alpha_1,\alpha_2}$ hereafter, shares the high incompressibility of the QMC functional. However, this parameterisation does not perform well, yielding binding energies approximately 60 MeV away from experiment for tin nuclei, as shown in Fig.~\ref{fig:SQMC4_Sn}, due to the high incompressibility.
\par
The incompressibility of the QMC is expected to be reduced by the inclusion of Fock terms involving pion exchange, reducing the incompressibility to be within the bounds recently deduced from an extensive review of experimental data in Ref.~\cite{stone2014} of 250~--~315~MeV.  Indeed, the relativistic version~\cite{stone2007} of the QMC EDF, including pion exchange, lowers the incompressibility from 340~MeV to $\sim300$~MeV.
\par
As a non-relativistic version of the QMC model with finite range terms coming from pion exchange remains to be developed, and the high-incompressibility functional SQMC$^{\alpha_1,\alpha_2}$ performs poorly, we will focus on QMC700, QMC700* and SQMC, where the restricted density dependence gives a low incompressibility and substantially more accurate results.
%
\section{\label{sec:Results}Isovector effects}
In the following, the isovector dependence of the SQMC functional will be investigated by comparing its behaviour to that of QMC700*, as the latter employs the same treatment of the density dependence, only differing in its exclusion of the four parameters $\{x_{1-3},W_0'\}$ in the fitting procedure.
\par
Hartree-Fock-Bogoliubov codes are a class of microscopic mean-field approach which are well-suited to the study of exotic nuclear structure, as the treatment of pairing remains robust for weakly bound nuclei. 
All of the results that follow were calculated using the \textsc{hfbtho}v2.00d code \cite{HFBTHO}, which allows for axial deformation. \textsc{hfbtho} also has all the properties required to properly implement the modern Universal Nuclear Energy Density Functional~1 (UNEDF1) \cite{Kortelainen2012} Skyrme parameterisation. UNEDF1 is used below as a representative of the current generation of phenomenological functionals and their level of accuracy.
\par
Long isotopic and isotonic chains provide a large range of differences between proton and neutron number, useful for revealing trends in the isovector properties of the functionals. HFB calculations for the tin ($Z=50$) and lead ($Z=82$) isotopic chains, and the $N=126$ isotonic chain give a systematic picture of each functional's performance. We compare our various theoretical calculations to the experimental binding energies reported in the 2012 Atomic Mass Evaluation (AME2012 \cite{AME2012}), labelled $E_{\textrm{exp}}$ in what follows. 
\par
All calculations were performed without any centre-of-mass corrections. Indeed the parameters of UNEDF1 were fitted without these corrections so that it can be used both in structure calculations and in time-dependent Hartree-Fock simulations of heavy ion collisions \cite{Simenel2012,Kortelainen2012}.
\par
The mixed (surface and volume) pairing effective interaction takes the form
\begin{equation}
V_{\textrm{pair}}(\bm{r},\bm{r'}) = V_{0}\left(1-\frac{1}{2}\frac{\rho(\bm{r})}{\rho_{0}}\right)\delta\left(\bm{r}-\bm{r}'\right),
\end{equation}
in the $^1S_0$ channel with the pairing strength $V_{0}$ adjusted for each parameterisation to give a neutron pairing gap of 1.245 MeV for $^{120}$Sn. UNEDF1 calculations use the proton and neutron pairing strengths prescribed in Ref.~\cite{Kortelainen2012}. All calculations include the Lipkin-Nogami (LN) prescription for particle-number projection~\cite{LN}.
\par
\begin{figure}
\includegraphics[width=\columnwidth]{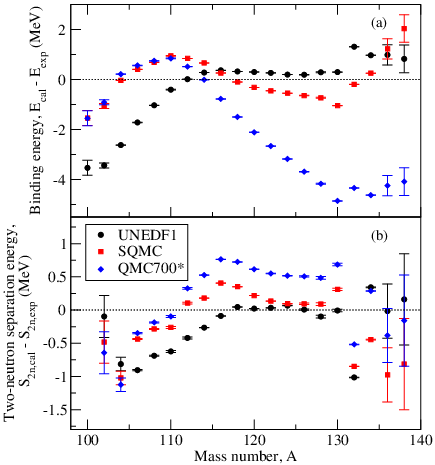}
\caption{\label{fig:Sn_combined}Difference between calculated and experimental energies for $^{A}$Sn nuclei: (a) ground-state binding energies, and (b) two-neutron separation energies. The error bars are from the experimental data $(E_{\textrm{exp}})$ \cite{AME2012}.}
\end{figure}
\begin{figure}
\includegraphics[width=\columnwidth]{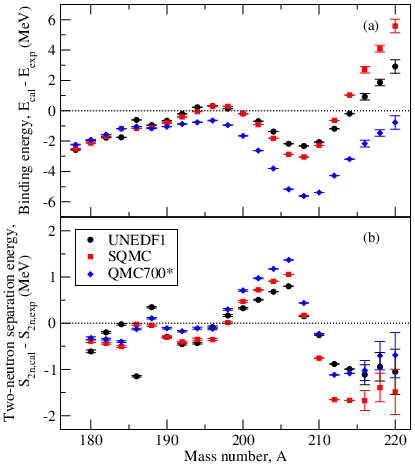}
\caption{\label{fig:Pb_combined}Same as Fig.~\ref{fig:Sn_combined} for $^A$Pb nuclei.}
\end{figure}
\begin{figure}
\includegraphics[width=\columnwidth]{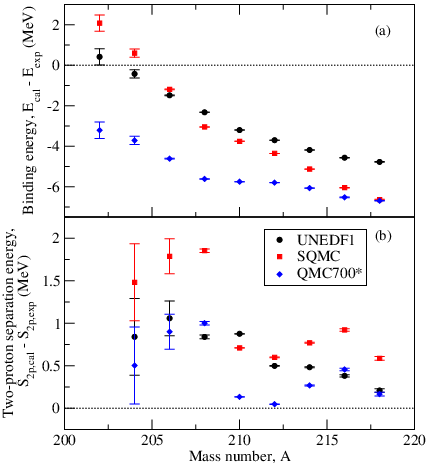}
\caption{\label{fig:N126_combined}(a) Same as Fig.~\ref{fig:Sn_combined}(a) for the $N=126$ isotonic chain, and (b) two-proton separation energies.}
\end{figure}

Figure~\ref{fig:Sn_combined} shows the ground-state binding energies for the experimentally measured even-even tin nuclei as calculated using the UNEDF1, SQMC and QMC700* parameterisations. Figures~\ref{fig:Pb_combined} and \ref{fig:N126_combined} repeat this analysis for the lead isotopes and for the $N=126$ isotonic chain, respectively.
\par
Comparing the results of QMC700* to SQMC it becomes apparent that while a single parameter ($x_{0}$) controlling the isovector dependence is adequate for nuclei close to symmetry, the extra terms included in SQMC quickly become significant for neutron-rich systems. Except for a handful of nuclei, the binding energies given by SQMC are significantly closer to experiment than those of QMC700*, supporting the isovector dependence microscopically derived within the QMC model.
\par
The accuracy of the SQMC results is comparable to that of UNEDF1, especially away from stability. This is particularly remarkable when one recalls that no experimental masses were directly used in the production of the SQMC, only nuclear matter pseudodata and the QMC model. UNEDF1 on the other hand, included the masses of $^{108, 112-124}$Sn and $^{198-214}$Pb in its fitting procedure \cite{Kortelainen2010}.
\par
Also shown in Figs.~\ref{fig:Sn_combined} and \ref{fig:Pb_combined} are the two-neutron separation energies, $S_{\textrm{2n}}$, for the same nuclei, where
\begin{equation}
S_{\textrm{2n, cal}}\left(N, Z\right) = E_{\textrm{cal}}\left(N-2, Z\right) - E_{\textrm{cal}}\left(N, Z\right).
\end{equation}
Similarly, Fig.~\ref{fig:N126_combined} shows the two-proton separation energies. While the trends are less clear for these separation energies than for the total binding energies, all three parameterisations perform well, generally lying within 2 MeV of the experimental data.

\section{\label{sec:Spin-orbit}The spin-orbit functional}
The nuclear spin-orbit interaction is a manifestation of relativity. Non-relativistic models, such as the Skyrme interaction, must add it by hand. The isovector properties of the spin-orbit interaction have remained largely unknown, with little theoretical guidance available and significant difficulties faced extracting them from experimental data~\cite{Bender2003}.

\subsection{Isovector dependence}
In the case of the Skyrme functional, the isovector dependence of the spin-orbit functional is controlled by the relationship between the two terms of Eq.~(\ref{eq:HsoSkyrme}), i.e. the ratio of the parameters $W_{0}'$ to $W_{0}$. This ratio offers a convenient way to compare the predictions of different approaches, as done in Table~\ref{tab:W0'/W0}.
 \begin{table}
 \caption{\label{tab:W0'/W0}Ratio of the spin-orbit coupling constants, illustrating the isovector dependence of spin-orbit energy density functionals.}
 \begin{ruledtabular}
 \begin{tabular}{l l}
   Model & $W_{0}'/W_{0}$ \\
\colrule
   Hartree & 0 \\
   Hartree-Fock (Standard Skyrme) & 1 \\
   UNEDF1 (Modern Skyrme) & 1.86 \\
   (S)QMC & 1.78 \\
    & \\
   Standard RMF \cite{Sharma1995} & $\sim 0.1$ \\
   (S)QMC, Hartree terms only, Dirac $\mu$ & 0.2
 \end{tabular}
 \end{ruledtabular}
 \end{table}
\par
When only Hartree terms are considered the zero-range, two-body spin-orbit force of the Skyrme interaction necessitates that the resulting functional have only an isoscalar term, with $W_{0}'=0$. The inclusion of the Fock (exchange) term introduces an isovector dependence, however there remains only one spin-orbit parameter, as $W_{0}'$ is fixed equal to $W_{0}$. While this is sufficient for a good fit near stability, a single degree-of-freedom allows for no control over the isovector dependence and no guarantee of reliability when extrapolating to exotic nuclei.
\par
By relaxing the connection to the two-body Skyrme interaction, it is possible to introduce a second free spin-orbit parameter at the level of the energy density functional. Though it was first proposed by Reinhard and Flocard in 1995 \cite{Reinhard1995}, it was not widely utilised until the recent UNEDF parameterisations \cite{Kortelainen2010, Kortelainen2012,kortelainen2014}. An extensive fit to medium and heavy mass nuclei yielded a significantly stronger isovector dependence for UNEDF1 $\left(W_{0}'/W_{0}=1.86\right)$ than standard Skyrme forces.
\par
In the QMC model the nucleons are assumed to move slowly and non-relativistic limits are taken to obtain the energy density functional. However, as the model is based upon a relativistic model of quarks and mesons, important relativistic corrections remain, chief among them the spin-orbit interaction. The spin-orbit interaction derived within the QMC model emerges from a combination of the variation of the vector meson fields $\left(\omega\textrm{ and }\rho\right)$ across the finite nucleon, and Thomas precession, a purely relativistic effect induced when changing frames of reference. The resulting density functional [Eq.~(\ref{eq:HsoQMC})] has no extra free parameters, unlike non-relativistic approaches.
\par
The microscopically derived isovector dependence of the (S)QMC functional is remarkably similar to that of the phenomenological UNEDF1 functional (see Tables~\ref{tab:Parameters} and \ref{tab:W0'/W0}).
The significance of the similarity between the spin-orbit functionals of the QMC model and UNEDF1 may be determined by examining whether the isovector dependence of the spin-orbit term has any impact on Hartree-Fock calculations of nuclei. The spin-orbit terms of the SQMC parameterisation and the QMC model are identical, making SQMC a convenient tool to test the isovector dependence of the original QMC functional.
\par
\begin{figure}
\includegraphics[width=\columnwidth]{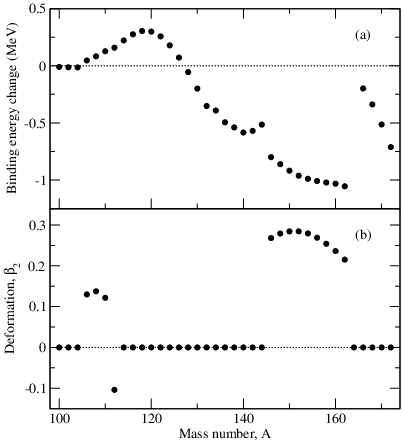}
\caption{\label{fig:EBchange_Sn}(a) Binding energy difference between SQMC with $W'_{0}/W_{0}=1.78$ and SQMC with $W'_{0}/W_{0}=1$, for $^{A}$Sn nuclei. (b) Quadrupole deformation, $\beta_{2}$, predicted by the SQMC parameterisation.} 
\end{figure}
\begin{figure}
\includegraphics[width=\columnwidth]{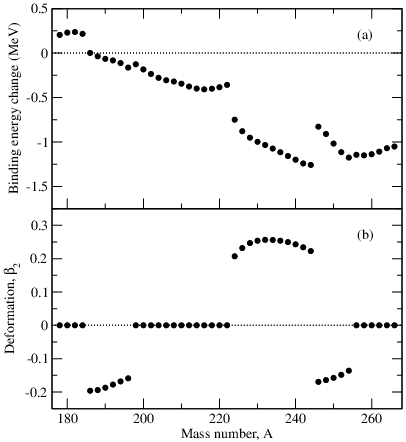}
\caption{\label{fig:EBchange_Pb}Same as Fig.~\ref{fig:EBchange_Sn} for $^{A}$Pb nuclei.}
\end{figure}
\begin{figure}
\includegraphics[width=\columnwidth]{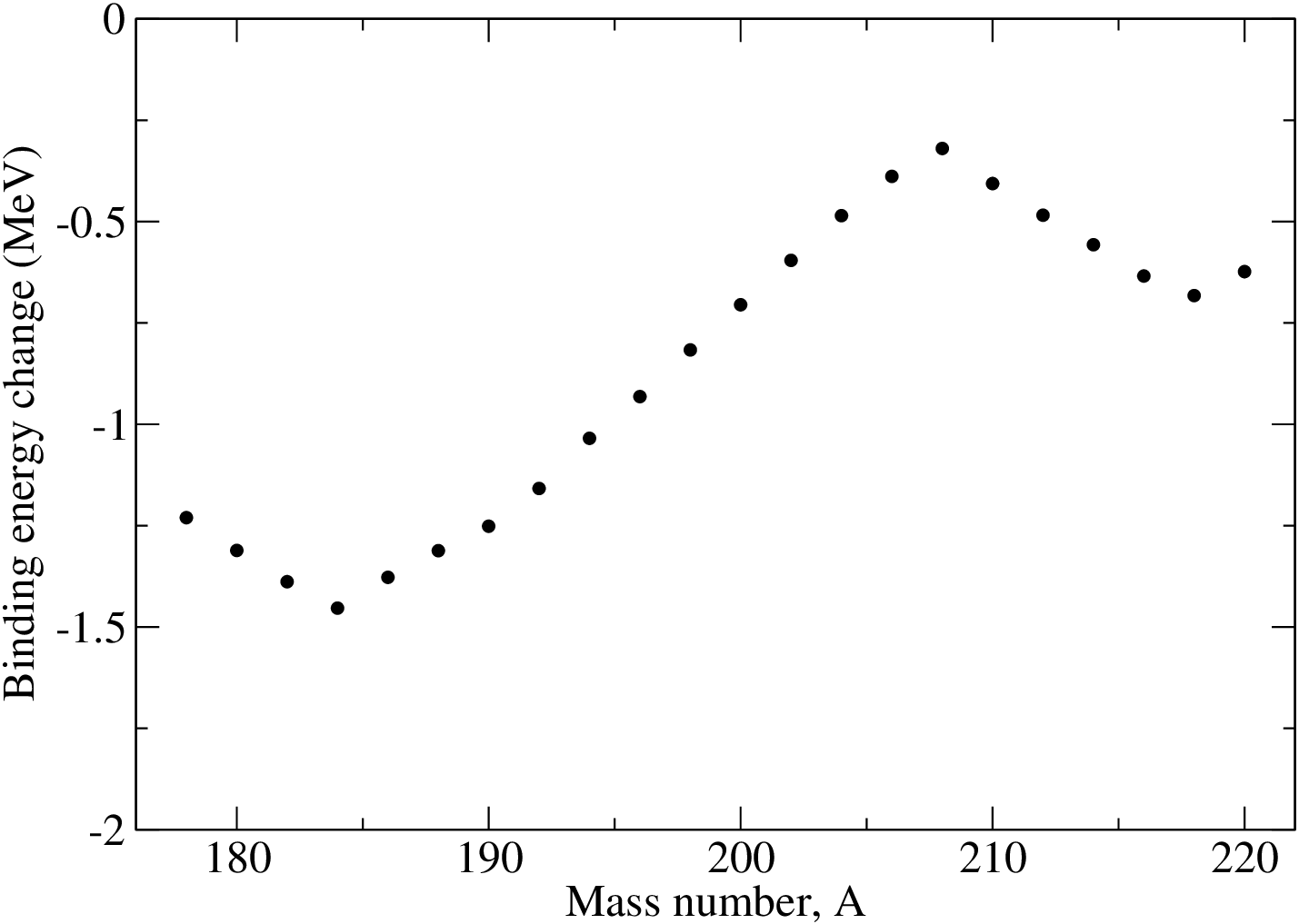}
\caption{\label{fig:EBchange_N126}Same as Fig.~\ref{fig:EBchange_Sn}(a) for $N=126$ isotonic chain.}
\end{figure}

It is possible to extract the impact of only the isovector dependence of the spin-orbit term by comparing the SQMC parameterisation (with $W_{0}'/W_{0}=1.78$) against a baseline which has only one spin-orbit parameter $\left(W_{0}'=W_{0}=104.3872\mbox{ MeV fm}^5\right)$ but is otherwise identical. \textsc{hfbtho} calculations, using the same pairing strength for both parameterisations, were performed for all bound nuclei of the Sn and Pb isotopic chains and the $N=126$ isotonic chain, and are reported in Figures~\ref{fig:EBchange_Sn}, \ref{fig:EBchange_Pb} and \ref{fig:EBchange_N126}, respectively.
\par
The figures show a relatively strong isovector dependence, varying rapidly with $N$ and $Z$. 
The isovector dependence of the spin-orbit term induces a binding energy change of approximately 1 MeV in the regions of the r-process ($^{134 - 152}$Sn, $^{238 - 266}$Pb and $^{180 - 184}Z_{126}$) as predicted by SQMC. A change of only 500 keV for nuclei in the region of $^{140}$Sn is already expected to have a significant impact on r-process abundances \cite{Mumpower2016}. This illustrates that it is crucial to properly account for the isovector properties of the spin-orbit functional, as included in UNEDF1 and predicted in the QMC model.

\subsection{\label{sec:Def}Impact of deformation}

The lower panels of Figs.~\ref{fig:EBchange_Sn} and \ref{fig:EBchange_Pb} show the deformation of the mean-field ground-states in tin and lead isotopes. 
We observe that the effect of the isovector contribution to the spin-orbit functional is clearly enhanced by the presence of prolate deformation in neutron-rich nuclei ($^{146 - 162}$Sn and $^{224 - 252}$Pb).  
A more detailed investigation of the interplay between deformation and the isovector contribution to the spin-orbit functional will be the focus of a future study.
\begin{figure}
\includegraphics[width=\columnwidth]{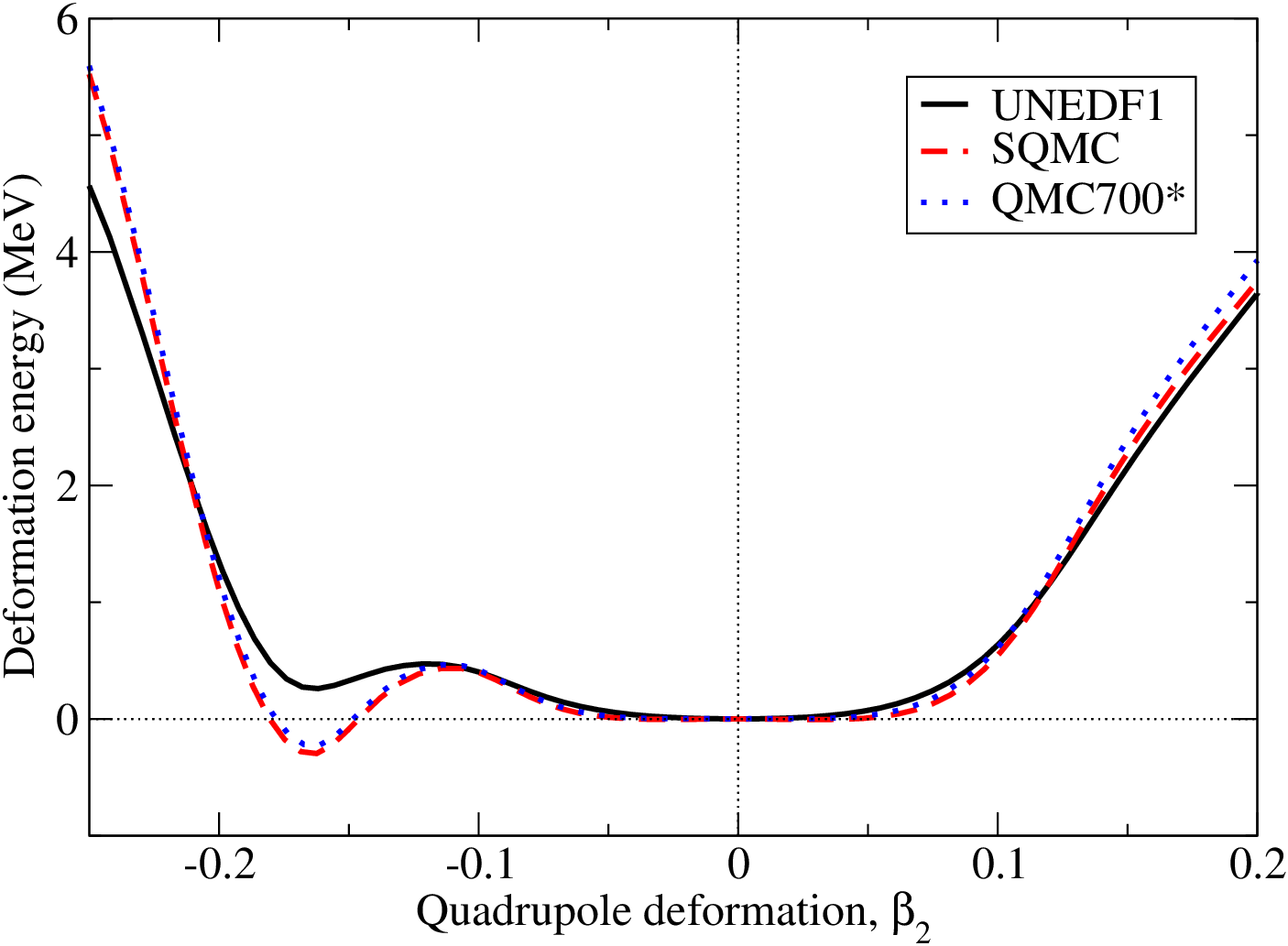}
\caption{\label{fig:def196}Mean-field deformation energy as a function of the quadrupole deformation parameter $\beta_2$ in $^{196}$Pb.}
\end{figure}
\begin{figure}
\includegraphics[width=\columnwidth]{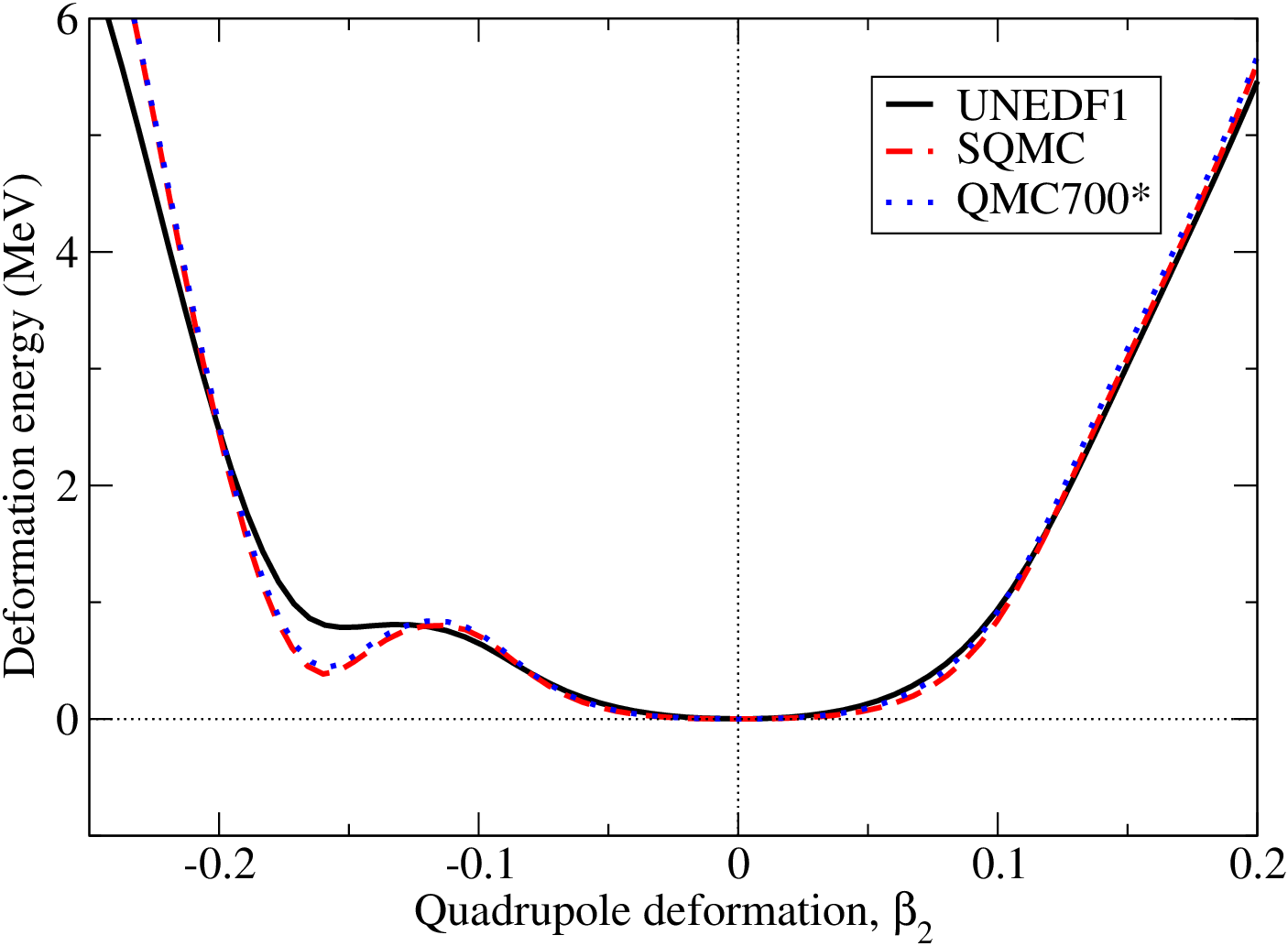}
\caption{\label{fig:def198}Same as Fig.~\ref{fig:def196} for $^{198}$Pb.}
\end{figure}
We should point out that the deformations reported in Figs.~\ref{fig:EBchange_Sn} and \ref{fig:EBchange_Pb} are those of the lowest energy mean-field state in the potential energy curve.
In reality, the ground-state of soft nuclei is expected to be made of a superposition of mean-field states with various deformations. 
This is illustrated in Figs.~\ref{fig:def196} and~\ref{fig:def198} showing the potential energy versus the quadrupole deformation parameter $\beta_2$ in $^{196}$Pb and $^{198}$Pb nuclei, respectively, for UNEDF1, QMC700* and SQMC parameterisations. 
Interestingly, QMC700* and SQMC both predict an oblate minimum in $^{196}$Pb while the UNEDF1 minimum is spherical. 
In $^{198}$Pb, all functionals predict a spherical minimum. 
However, all parameterisations predict a soft potential with respect to quadrupole deformation. 
An improved description of the ground-states in such nuclei would thus require beyond mean-field approaches such as the generator coordinate method (GCM) \cite{Niksic2006,Bender2008,Rodriguez2010,Yao2010}.
\subsection{\label{sec:RMF}Comparison to relativistic mean-field}
The term ``relativistic mean-field'' (RMF) encompasses a constantly expanding set of theories. The version discussed here is based on a hadronic Lagrangian of inert nucleons exchanging $\sigma$, $\omega$ and $\rho$ mesons \cite{Sharma1995}. Like the QMC model, RMF is able to predict a nuclear spin-orbit interaction. However, it is often argued that, in RMF, Fock terms are negligible \cite{Bender2003} and Dirac nucleon magnetic moments, $\mu_{p}=1$ and $\mu_{n}=0$, are used.
\par
Once again we use the ratio $W_0'/W_0$ to compare the isovector dependence of the spin-orbit energy density functionals in Table~\ref{tab:W0'/W0}. RMF gives a substantially weaker isovector dependence than the QMC model, UNEDF1 and even standard Skyrme. RMF is very close to the Hartree value of $W_{0}'=0$, only deviating due to a small contribution from the vector-isovector $\rho$ meson.
\par
By keeping only the direct (Hartree) terms of the QMC spin-orbit functional and setting the nucleon magnetic moments to their Dirac rather than the experimental ($\mu_{s}=\mu_{p}+\mu_{n}=0.88$, $\mu_{v}=\mu_{p}-\mu_{n}=4.7$) values, one would obtain a similarly weak isovector dependence (Tab.~\ref{tab:W0'/W0}). This implies that the weak isovector dependence of RMF is primarily due to these two approximations: neglect of exchange terms and use of Dirac magnetic moments. These are approximations which are not required in the QMC model.
%
\section{\label{sec:Conc}Conclusion}
The predictions of ground-state masses from the SQMC parameterisation are of a comparable level of accuracy to UNEDF1 for the studied systems, showing the promise of replacing some level of phenomenology with input from a more microscopic theory based on quark degrees-of-freedom. The isovector behaviour of the functional can be derived from the QMC model and is significant for exotic nuclei. It is remarkably similar to that of UNEDF1 and much stronger than RMF. This is significant because this dependence is an important factor, which must be taken into account in r-process calculations.
\par
The SQMC functional may be used in any existing Skyrme-Hartree-Fock codes to study further exotic structure effects such as shell evolution, driplines or superheavy magic numbers. Also, using a time-dependent Hartree-Fock code, dynamic processes can be investigated, including reactions with exotic nuclei, fusion barriers, transfer and fission, revealing how they are affected by the high-energy physics upon which the QMC model is based.
\par
It will also be possible to perform similar investigations using the original QMC functional, rather than the SQMC parameterisation, by modifying Skyrme-based Hartree-Fock codes to accept the more complicated density dependence of the central terms. Stone \textit{et al.} \cite{Stone2016} modified the static \textsc{skyax} code \cite{skyax} in this way and used nuclear data to fix the few free parameters of the QMC model. They found its performance across the nuclear chart to be on a level comparable to a traditional Skyrme functional with many more parameters and particularly impressive for the masses of superheavy nuclei not included in the fit. A similar implementation in a time-dependent code will allow for a study of the giant monopole resonance to illuminate the issue of nuclear matter incompressibility and the importance of single pion exchange, among many other dynamic processes.

\begin{acknowledgments}
CS thanks G. Lane for useful discussions. 
This research was undertaken with the assistance of resources from the National Computational Infrastructure (NCI), which is supported by the Australian Government. 
This research was supported by an Australian Government Research Training Program (RTP) Scholarship and by the Australian Research Council through ARC Grants FT120100760 (CS), DP160101254 (CS), FL110100098 (ECS), DP170102423 (ECS), DP150103164 (AWT) and by the ARC Centre of Excellence for Particle Physics at the Terascale, CE110001104 (AWT). 
\end{acknowledgments}


\bibliography{PaperRefs}

\end{document}